\documentclass[epj]{svjour}
\usepackage{graphics}
\usepackage[fleqn]{amsmath}
\usepackage{appendix}
\begin{document}

\title{Statistical model analysis of hadron yields in proton-nucleus and heavy-ion collisions at SIS 18 energies}
%\subtitle{}
\author{G.~Agakishiev$^{7}$,O.~Arnold$^{10,9}$, A.~Balanda$^{3}$, D.~Belver$^{18}$, A.~Belyaev$^{7}$,
J.C.~Berger-Chen$^{10,9}$, A.~Blanco$^{2}$, M.~B\"{o}hmer$^{10}$, J.~L.~Boyard$^{16}$, P.~Cabanelas$^{18}$
% ...
}

%E.~Castro$^{17}$, M.~Destefanis$^{11}$,F.~Dohrmann$^{6}$,  C.~Gilardi$^{11}$, J.~S.~Lange$^{11}$,  E.~Morini\`{e}re$^{15}$, J.~Mousa$^{14}$,L.~Rehnisch$^{8}$,T.~Scheib$^{8}$,H.~Schuldes$^{8}$,  M.~Wisniowski$^{3}$,

\institute{(HADES collaboration) \\
\mbox{$^{1}$Istituto Nazionale di Fisica Nucleare - Laboratori Nazionali del Sud, 95125~Catania, Italy}\\
\mbox{$^{2}$LIP-Laborat\'{o}rio de Instrumenta\c{c}\~{a}o e F\'{\i}sica Experimental de Part\'{\i}culas , 3004-516~Coimbra, Portugal}\\
\mbox{$^{3}$Smoluchowski Institute of Physics, Jagiellonian University of Cracow, 30-059~Krak\'{o}w, Poland}\\
\mbox{$^{4}$GSI Helmholtzzentrum f\"{u}r Schwerionenforschung GmbH, 64291~Darmstadt, Germany}\\
\mbox{$^{5}$Technische Universit\"{a}t Darmstadt, 64289~Darmstadt, Germany}\\
\mbox{$^{6}$Institut f\"{u}r Strahlenphysik, Helmholtz-Zentrum Dresden-Rossendorf, 01314~Dresden, Germany}\\
\mbox{$^{7}$Joint Institute of Nuclear Research, 141980~Dubna, Russia}\\
\mbox{$^{8}$Institut f\"{u}r Kernphysik, Goethe-Universit\"{a}t, 60438 ~Frankfurt, Germany}\\
\mbox{$^{9}$Excellence Cluster 'Origin and Structure of the Universe' , 85748~Garching, Germany}\\
\mbox{$^{10}$Physik Department E12, Technische Universit\"{a}t M\"{u}nchen, 85748~Garching, Germany}\\
\mbox{$^{11}$II.Physikalisches Institut, Justus Liebig Universit\"{a}t Giessen, 35392~Giessen, Germany}\\
\mbox{$^{12}$Istituto Nazionale di Fisica Nucleare, Sezione di Milano, 20133~Milano, Italy}\\
\mbox{$^{13}$Institute for Nuclear Research, Russian Academy of Science, 117312~Moscow, Russia}\\
\mbox{$^{14}$Institute of Theoretical and Experimental Physics, 117218~Moscow, Russia}\\
\mbox{$^{15}$Department of Physics, University of Cyprus, 1678~Nicosia, Cyprus}\\
\mbox{$^{16}$Institut de Physique Nucl\'{e}aire (UMR 8608), CNRS/IN2P3 - Universit\'{e} Paris Sud, F-91406~Orsay Cedex, France}\\
\mbox{$^{17}$Nuclear Physics Institute, Academy of Sciences of Czech Republic, 25068~Rez, Czech Republic}\\
\mbox{$^{18}$LabCAF. F. F\'{\i}sica, Univ. de Santiago de Compostela, 15706~Santiago de Compostela, Spain}\\
\\
\mbox{$^{a}$ also at Lawrence Berkeley National Laboratory, ~Berkeley, USA}\\
\mbox{$^{b}$ also at ISEC Coimbra, ~Coimbra, Portugal}\\
\mbox{$^{c}$ also at ExtreMe Matter Institute EMMI, 64291~Darmstadt, Germany}\\
\mbox{$^{d}$ also at Technische Universit\"{a}t Dresden, 01062~Dresden, Germany}\\
\mbox{$^{e}$ also at Dipartimento di Fisica, Universit\`{a} di Milano, 20133~Milano, Italy}\\
\mbox{$^{f}$ also at Frederick University, 1036~Nicosia, Cyprus}\\
\mbox{$^{g}$ also at Dipartimento di Fisica and INFN, Universit\`{a} di Torino, 10125~Torino, Italy}\\
\mbox{$^{h}$ also at Utrecht University, 3584 CC~Utrecht, The Netherlands}\\
\\
\mbox{$^{\ast}$ corresponding author: m.lorenz@gsi.de}
}
\date{Received: 22.12.2015 / Revised version: date}

\abstract{
The HADES data from p+Nb collisions at center of mass energy of $\sqrt{s_{NN}}$= 3.2 GeV are analyzed by employing a statistical model. Accounting for the identified hadrons $\pi^0$, $\eta$, $\Lambda$, $K^{0}_{s}$, $\omega$ allows a surprisingly good description of their abundances with parameters $T_{chem}=(99\pm11)$ MeV and $\mu_{b}=(619\pm34)$ MeV, which fits well in the chemical freeze-out systematics found in heavy-ion collisions. In supplement we reanalyze our previous HADES data from Ar+KCl collisions at $\sqrt{s_{NN}}$= 2.6 GeV with an updated version of the statistical model. We address equilibration in heavy-ion collisions by testing two aspects: the description of yields and the regularity of freeze-out parameters from a  statistical model fit. Special emphasis is put on feed-down contributions from higher-lying resonance states which have been proposed to explain the experimentally observed $\Xi^-$ excess present in both data samples. 
}

%\authorrunning

%\titlerunning
\authorrunning{The HADES collaboration (G.~Agakishiev {\it et al.})}
\titlerunning{Statistical model analysis of hadron multiplicities at SIS 18 energy}
\maketitle

\section{Introduction}
\label{intro}
The idea of applying statistical methods to predict hadron yields in collisions of ions goes back to Heinz Koppe in 1948 \cite{koppe} as recently pointed out in \cite{tawfik}.  Half a century later statistical hadronization models have been established as a successful tool to describe particle yields or yield ratios from relativistic and ultrarelativistic heavy-ion collisions (HICs) \cite{BraunMunzinger:2003zd,Tawfik:2014eba,Floris} with only a few parameters. Moreover, the extracted freeze-out parameters show a striking regularity, lining up on a curve in the temperature - baryochemical potential plane, connecting smoothly data from the lowest energies at SIS18 up to the highest available energy at LHC \cite{Cleymans:2005xv}.\\ 
These findings give a strong hint that the observed inclusive, ensemble-averaged hadron abundances correspond to (but need not to be identical with) a state described by thermal and chemical equilibrium.\\
Since the days of Hagedorn \cite{hage}, statistical methods have also been used to predict particle production in elementary reactions, see e.g.  \cite{shuryak}. More recently a detailed analysis applying exactly the same model \cite{Becattini:2003wp}, which successfully describes hadron yields in HICs, shows also a good agreement for yields and even transverse momentum spectra obtained in elementary $e^{+}+e^{-}$ and $p+p$ collisions \cite{Becattini:2001fg,Becattini:2008tx}. \footnote{For completeness we refer the reader to another recent statistical analysis reaching different conclusions \cite{andronic_el}.} These findings question conclusions drawn about chemical equilibrium (either instantaneously or as time projection) in heavy-ion collisions based on the comparison of data to hadron yields obtained via statistical model calculations and ask for a more fundamental reason for the good agreement. \\
In this context it is important to discuss the distinctions between the different realizations of statistical models, especially their treatment of non-equilibrium parameters. While in \cite{Andronic:2005yp,Stachel2013} a grand canonical ensemble with only the parameters $T$ (temperature) and $\mu_{B}$ (baryo-chemical potential) is used for central heavy-ion collisions, the authors of \cite{Becattini:2003wp} are using a mixed canonical ensemble, conserving strangeness exactly plus an additional multiplicative factor $\gamma_{s}$ in order to additionally suppress particles containing strangeness. In \cite{Cleymans:1999,Kraus:2007hf} the authors use also a mixed canonical ensemble but introduce a strangeness correlation volume parameter $V_c$ (or correlation radius parameter $R_c$) instead of $\gamma_{s}$.  In \cite{Petran}, on the other hand, $\gamma_{s}$ plus an additional parameter, suppressing the light quarks u,d, called $\gamma_{q}$ is used.\\
The system size and centrality dependence of those non-equilibrium parameters have been investigated in \cite{cley1,cley2,bk3}. The authors find a significant increase of the strangeness suppression factor $\gamma_{s}$ with increasing system size.\\
We state that in our previous paper \cite{Agakishiev:2010rs}, applying a thermal fit to hadron yields obtained from Ar+KCl reactions at 1.76A GeV, we find the necessity for an additional volume parameter $V_c$ ($R_c$) to further suppress strangeness and to reproduce the single-strange particles. However, the double strange $\Xi^-$ hyperon yield overshoots the thermal fit by more than an order of magnitude.
Recently, feed down from higher-lying resonances has been proposed as a possible explanation for the observed  $\Xi^-$ excess \cite{Steinheimer:2015sha}. 
In addition, the hadron spectrum included in the statistical model THERMUS (v2.3) \cite{Wheaton:2004qb}, which we used in \cite{Agakishiev:2010rs} for our Ar+KCl data, has been updated according to the report of the particle data group (PDG) 2014 \cite{Agashe:2014kda} recently in THERMUS (v3.0).\\
In order to address the aspect of equilibration, we test here two aspects: the description of yields and regularity of freeze-out parameters by confronting data sets from p+Nb and Ar+KCl using the same statistical model and the same parameters. 
The statistical analysis of p+A represents the first of its kind in this energy regime, where usually the available yields of different particle species are limited. The HADES data allow for the first time a simultaneous fit to eight different measured yields in one experimental run.   
Special emphasis is put on the effect of the new states in the hadron spectrum, e.g. feed down, included in the PDG report in the last decade.\\
This paper is organized as follows:\\
We start with an overview of the two data samples in section 2, before we present and discuss the results of the statistical model fits in 3.1 and 3.2. Section 3.3 is devoted to the $\Xi^-$ excess. Our summary is given in section 4. Finally, in the appendix we include a discussion of statistical strangeness production at high baryochemical potential using exact strangeness conservation. 

\section{Data sample}
HADES is a charged-particle detector consisting of a 6-coil toroidal magnet centered around the beam axis and six identical detection sections located
between the coils and covering polar angles between $18^{\circ}$ and $85^{\circ}$.  Each sector is equipped with a Ring-Imaging Cherenkov (RICH)
detector followed by Multi-wire Drift Chambers (MDCs), two in front of and two behind the magnetic field, as well as a scintillator hodoscope (TOF/TOFino).  Hadron identification is based on the time-of-flight and on the energy-loss information from TOF/TOFino, as well as from the MDC tracking chambers. A detailed description of HADES is given in \cite{Agakishiev:2009am}.

\subsection{Ar+KCl at $\bf\sqrt{s_{NN}}$= 2.6 GeV}
An argon beam of $\sim 10^6$ particles/s was incident with a beam energy of 1.76A GeV on a four-fold segmented KCl target with a total thickness corresponding to $3.3$ $\%$ interaction probability. A fast diamond start detector located upstream of the target was intercepting the beam and was used to determine the time-zero information. The data readout was started by a first-level trigger (LVL1) requiring a charged-particle multiplicity, $MUL \ge 16$, in the scintillator hodoscope. About $7.4  \times 10^8$ LVL1 events have been collected.
The yields of the various identified particles obtained in \cite{Agakishiev:2010rs,schuldes_fair,Pavel,Agakishiev:2009ar,Agakishiev:2010zw,Agakishiev:2013nta,Agakishiev:2011vf,Agakishiev:2009rr} and their inverse slope parameter $T_{eff}$ obtained from fitting Boltzmann distributions to the transverse mass spectra at mid-rapidity are listed in Tab. \ref{tab_arkcl}. The value of $A_{part}$ is obtained by comparing the charged-particle multiplicity to the UrQMD transport model \cite{Agakishiev:2010rs,UrQMD}, while the yield of the $\eta$ meson is interpolated from TAPS measurements in Ca+Ca collisions at 1.5 and 2A GeV  \cite{Averbeck:2000sn}.  
The bias of the LVL2 trigger used to trigger on electrons and relevant for the $\omega$ is at the order of 10$\%$ and is corrected for. 

\begin{table*}
\caption{Multiplicities (i.e. yield/LVL1 event) and effective temperatures $T_{eff}$
  of particles produced in Ar(1.76A GeV)+KCl reactions. If only a single error is given, the value corresponds to the total error, including systematic and statistical uncertainties. A \textquotedblleft $-$\textquotedblright ~in the  $T_{eff}$ column means that the spectra are too scarce to extract a value.}
\label{tab_arkcl}       % Give a unique label

\begin{center}
\begin{tabular}{|c|c|c|c|}
    \hline
    Particle & Multiplicity & $T_{eff}$ [MeV] & Reference\\
    \hline
    \hline
    $\langle A_{part} \rangle$ & $38.5 \pm 4 $  & $ -$& \cite{Agakishiev:2010rs,UrQMD} \\
    \hline
    $p$ & $22.11 \pm 2.4 $  & $142 \pm 5$&  \cite{schuldes_fair}\\
    \hline
    $\pi^-$ & $3.9 \pm 0.19 \pm 0.34(syst)$  & $82.4 \pm 0.1 ^{+9.1}_{-4.6}$& \cite{Pavel} \\
   \hline
    $\eta$ & $0.081\pm 0.02$  & $-$ &  \cite{Averbeck:2000sn}\\
    \hline
    $\Lambda+\Sigma^0$ & $(4.09 \pm 0.1  \pm 0.17(extr) ^{+0.17}_{-0.37}(syst)) \times 10^{-2}$ & $95.5 \pm 0.7  +2.2$ & \cite{Agakishiev:2010rs} \\
    \hline
    $K^+$  & $(2.8 \pm 0.2 \pm 0.1(syst) \pm 0.1(extr)) \times 10^{-2}$ & $89 \pm 1 \pm 2$ & \cite{Agakishiev:2009ar} \\
    \hline
    $K^{0}_{S}$ & $(1.15 \pm 0.05 \pm 0.09(syst)) \times 10^{-2}$ & $92\pm 2$ & \cite{Agakishiev:2010zw} \\
    \hline
    $K^-$ & $(7.1 \pm 1.5 \pm 0.3 (syst) \pm 0.1(extr)) \times 10^{-4}$  & $69 \pm 2 \pm 4$& \cite{Agakishiev:2009ar} \\
    \hline
    $K^{*}(892)^0$ & $(4.4 \pm 1.1 \pm 0.5(syst) ) \times 10^{-4}$  & $-$ & \cite{Agakishiev:2013nta}\\
    \hline
    $\omega$ & $(6.7 \pm 2.7) \times 10^{-3}$  & $131\pm26$ & \cite{Agakishiev:2011vf} \\
    \hline
    $\phi$ & $(2.6 \pm 0.7 \pm 0.1 -0.3) \times 10^{-4}$  & $84 \pm 8$& \cite{Agakishiev:2009ar} \\
    \hline
    $\Xi^{-}$ & $(2.3 \pm 0.9) \times 10^{-4}$  & $-$ & \cite{Agakishiev:2009rr} \\
    \hline   
    \end{tabular}
\end{center}
\end{table*}

\subsection{p+Nb at $\bf\sqrt{s_{NN}}$= 3.2 GeV}
A proton beam of about $2 \times 10^6$ particles/s with kinetic energy of 3.5 GeV was incident
on a 12-fold segmented target of niobium ($^{93}$Nb). The first-level (LVL1) trigger required a charged-particle multiplicity $MUL \ge 3$ in the scintillator hodoscope.
About $3.2  \times 10^9$ LVL1 events have been collected. The yields of the various identified particles obtained in \cite{Agakishiev:2013noa,Agakishiev:2014kdy,Agakishiev:2014moo,xi_pNb} and their inverse slope parameter $T_{eff}$ obtained from fitting Boltzmann distributions to the transverse mass spectra are listed in Tab. \ref{tab_pNb}. The value for $A_{part}$ is obtained using a geometrical overlap model \cite{Agakishiev:2012vj,glauber}, while the 4$\pi$-yield of the $\omega$ meson is based on a GiBUU transport code which describes the data satisfactorily \cite{Agakishiev:2012vj,Weil:2012ji,Weil_priv}.  

\begin{table*}
\caption{As in Tab. 1 but for p(3.5GeV)+Nb reactions.}
\label{tab_pNb}       % Give a unique label

\begin{center}
\begin{tabular}{|c|c|c|c|}
    \hline
    Particle & Multiplicity & $T_{eff}$ [MeV] & Reference\\
    \hline
    \hline
    $\langle A_{part} \rangle$ & $2.8 \pm  0.6$  & $ -$& \cite{Agakishiev:2012vj,glauber}\\
    \hline
    $\pi^0$ & $0.66 \pm 0.06  \pm 0.1(syst)$  & $92\pm3$ combined fit with $\pi^-$& \cite{Agakishiev:2013noa} \\
   \hline
    $\pi^-$ & $0.6 \pm 0.1 $  & $92\pm3$ combined fit with $\pi^0$ &  \cite{Agakishiev:2013noa}  \\
   \hline
    $\eta$ & $0.034\pm 0.002 \pm 0.008(syst)$  & $84 \pm3$ &  \cite{Agakishiev:2013noa} \\
    \hline
    $\Lambda+\Sigma^0$ & $0.017 \pm 0.003 $ & $92 \pm 5$ & \cite{Agakishiev:2014kdy} \\
    \hline
    $K^{0}_{S}$ & $0.0055\pm0.0007$ &  $99\pm 4$ & \cite{Agakishiev:2014moo} \\
    \hline
    $\omega$ & $0.007\pm0.004$  & $-$ & \cite{Agakishiev:2012vj,Weil:2012ji,Weil_priv} \\
    \hline
    $\Xi^{-}$ & $(2.0 \pm 0.4 \pm0.6(syst)) \times 10^{-4}$  & $-$& \cite{xi_pNb} \\
    \hline
   
    \end{tabular}
\end{center}
\end{table*}

\section{Statistical model fit to hadron yields}
\subsection{Ar+KCl  at $\bf\sqrt{s_{NN}}$= 2.6 GeV}
\begin{figure}
\begin{center}
\resizebox{9cm}{!}{%
 \includegraphics{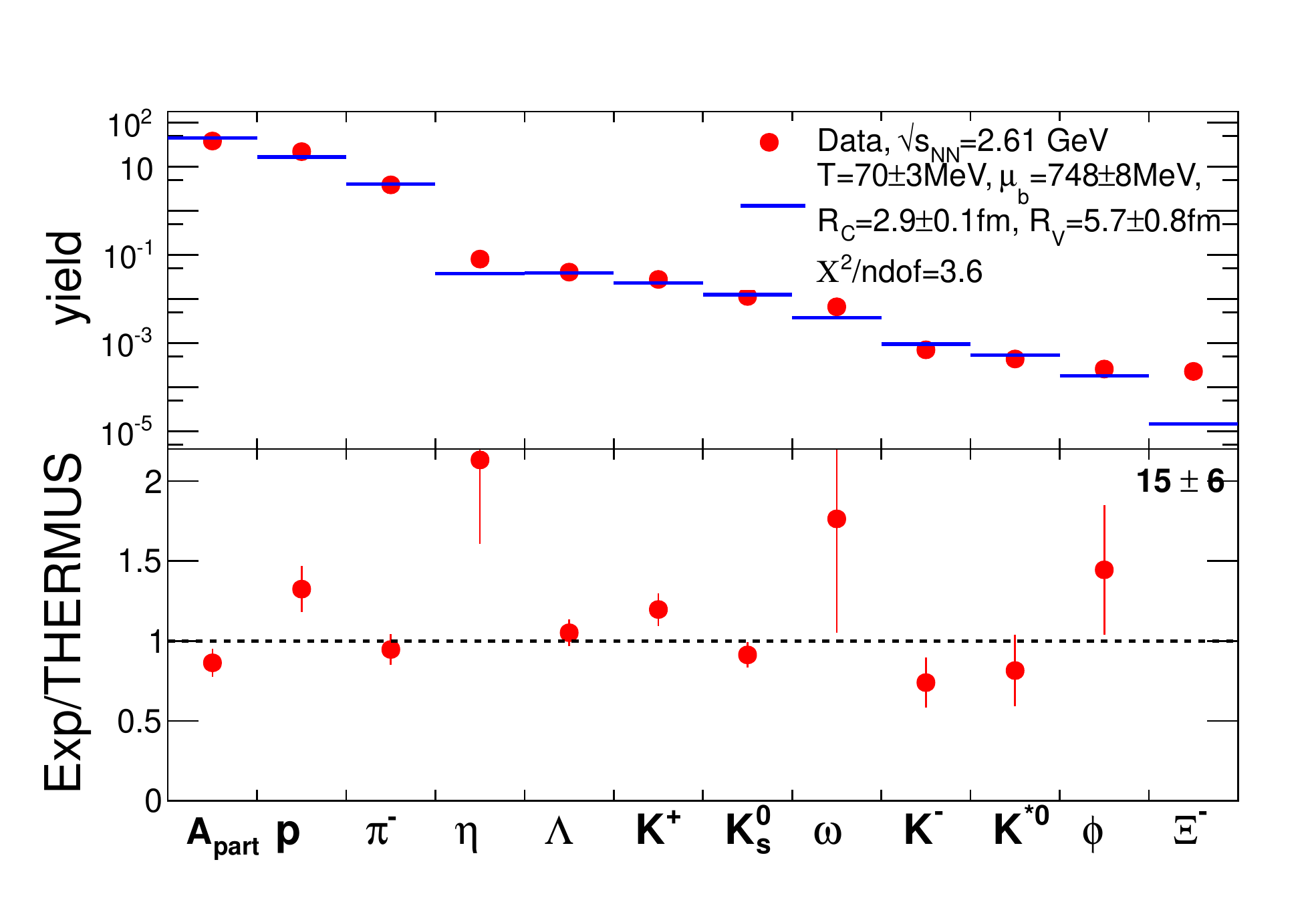}
% epja impact_parameter_herb.eps!!
}

\end{center}
%\vspace{5cm}       % Give the correct figure height in cm
\caption{Yields (filled red circles) of hadrons in Ar+KCl reactions and the corresponding THERMUS fit values (blue bars). The lower plot shows the ratio of the experimental value and the THERMUS value. For the $\Xi^{-}$ the ratio number is quoted instead of a point.}
\label{fig_arkcl}       % Give a unique label
\end{figure}

We apply a similar fit as in our previous work \cite{Agakishiev:2010rs} but use the updated version (v3.0) of THERMUS \cite{Wheaton:2004qb}.
The main difference to the previously used version (v2.3) is the included hadron spectrum which was updated from the PDG report 2002 \cite{pdg2002} to the one from 2014 \cite{Agashe:2014kda}, including now several new strange states, e.g. K*(800), as well as states containing charm, which are not relevant here.
In addition, we include now the experimental yields of the $p$, $\omega$ and $K^{*}(892)^0$ which have become available recently \cite{schuldes_fair,Agakishiev:2013nta,Agakishiev:2011vf}.\\
We use the mixed canonical ensemble where strangeness is exactly conserved while all other quantum numbers are calculated grand canonically and constrain the charge chemical potential $\mu_{Q}$ using the ratio of the baryon and charge numbers of the collision system. \\
The yield of the $\phi$ meson is of particular interest, because of its sensitivity to the strangeness suppression parameters $\gamma_{s}$ and $R_{c}$. As the $\phi$ conserves strangeness by definition as an $s\overline{s}$ state its yield is not suppressed in the $R_{c}$ formalism, while strongly suppressed when $\gamma_{s}$ is used.  We found in \cite{Agakishiev:2010rs} that the yield is well described using $R_{c}$ and therefore stick to this way of suppressing strange particle yields in our statistical model calculations. Note that at higher energies and small systems the description of the $\phi$ meson yield improves when additional suppression parameters are introduced \cite{Kraus:2007hf}.\\
We fit simultaneously all particle yields listed in Tab. \ref{tab_arkcl},  as well as the mean number of participants $\langle A_{part} \rangle$ and constrain the charge chemical potential $\mu_{Q}$. We find the following values for chemical freeze-out parameters $T_{chem}=(70\pm3)$ MeV, $\mu_{b}=(748\pm8)$ MeV, the strangeness correlation radius results as $R_{c}=(2.9\pm0.1)$ fm and the radius of the whole fireball $R=(5.7\pm0.8)$ fm with a $\chi^2$/d.o.f. of 3.6. 
A detailed comparison of the data with the statistical model fit is shown in the upper part of Fig. \ref{fig_arkcl}, while the lower part depicts the ratio of data to the THERMUS value. In case of the $\Xi^-$, a number is displayed instead of a data point. \\
The chemical composition of particles looks as it would have a single chemical freeze-out temperature $T_{chem}$. At collider energies this could be the hadronization temperature at the phase boundary of the quark-gluon-plasma and a hadron gas.\\
These values may be compared to the values of $T_{chem}=(76\pm2)$ MeV, $\mu_{b}=(799\pm22)$ MeV, $R_{c}=(2.2\pm0.2)$ fm, $R=(4.1\pm0.5)$ fm and $\chi^2$/d.o.f. of 2.6 obtained for the same system in \cite{Agakishiev:2010rs}.
We observe a deviation of all parameters at the order of a few standard deviations. 
While the percentaged deviation is only at the order of 5$\%$ for $T$ and $\mu_{b}$ it is about 25$\%$ for the radii. 
Due to correlations between all four parameters the minimum of the fit moves to a slightly different position in the parameter space as a result of an interplay of several effects. As more hadron states are included, the baryochemical potential is slightly lower, which is to some extent compensated by a larger volume. Due to this larger volume the temperature $T$ is slightly lower as otherwise the pion rate would overshoot the experimental values. \\
We state that we take the observed deviation of the freeze-out parameters to our previous work, as the expected systematic uncertainties of such fits. \\
Comparing the particle yields, the strongest deviations are observed for the protons, the $\eta$ and the $\Xi^-$. 
Already the results presented in \cite{Averbeck:2000sn} pointed out that the yield of the $\eta$ meson seems to favor a significantly higher freeze-out temperature. However, we want to point out that about half of its yield results of decays from baryon-resonances mainly $N(1535)$ in THERMUS. \\   
In the sector of the vector mesons both the yields of the $\omega$ and the $\phi$ are in favor of a slightly higher temperature, while the $K^{*}(892)^0$ yield is better described with a lower temperature. The latter observation is made also at higher energies and has been interpreted as a parameter for the lifetime of the hadronic phase within the chemical freeze-out at $T_{chem}$ and the kinetic freeze-out at $T_{kin}$. As due to the short life time of the $K^{*}(892)^0$ its decay products are rescattered inside the medium \cite{Markert:2002rw}, while for instance the decay products of the $\phi$ meson are not affected, as the $\phi$ decays mainly outside of the medium due to its longer life time. \\
The worse $\chi^2$/d.o.f. compared to the previous fit \cite{Agakishiev:2010rs} results mainly from the inclusion of the $\Xi^-$.
The excess of the experimentally measured $\Xi^-$ yield over the model decreases from a factor $24\pm9$ to a factor $15\pm6$ when using the actual version (v3.0) of THERMUS. We will come back to this in the discussion of the $\Xi^-$ excess.\\
The comparison of the extracted chemical freeze-out temperature $T_{chem}$ to the ones extracted from the inverse slope $T_{eff}$ of transverse mass spectra at mid-rapidity  for various particles listed in Tab. 1 is not straightforward. In a naive picture the extracted inverse slope parameter $T_{eff}$ include a pure kinetical component $T_{kin}$ plus an additive term, depending on the particle mass $m$ and the square of the radial expansion velocity $\beta$. In addition effects like resonance decays deform the spectra complicating this naive interpretation.
\subsection{p+Nb  at $\bf\sqrt{s_{NN}}$= 3.2 GeV}

\begin{figure}
\begin{center}
\resizebox{8cm}{!}{%
 \includegraphics{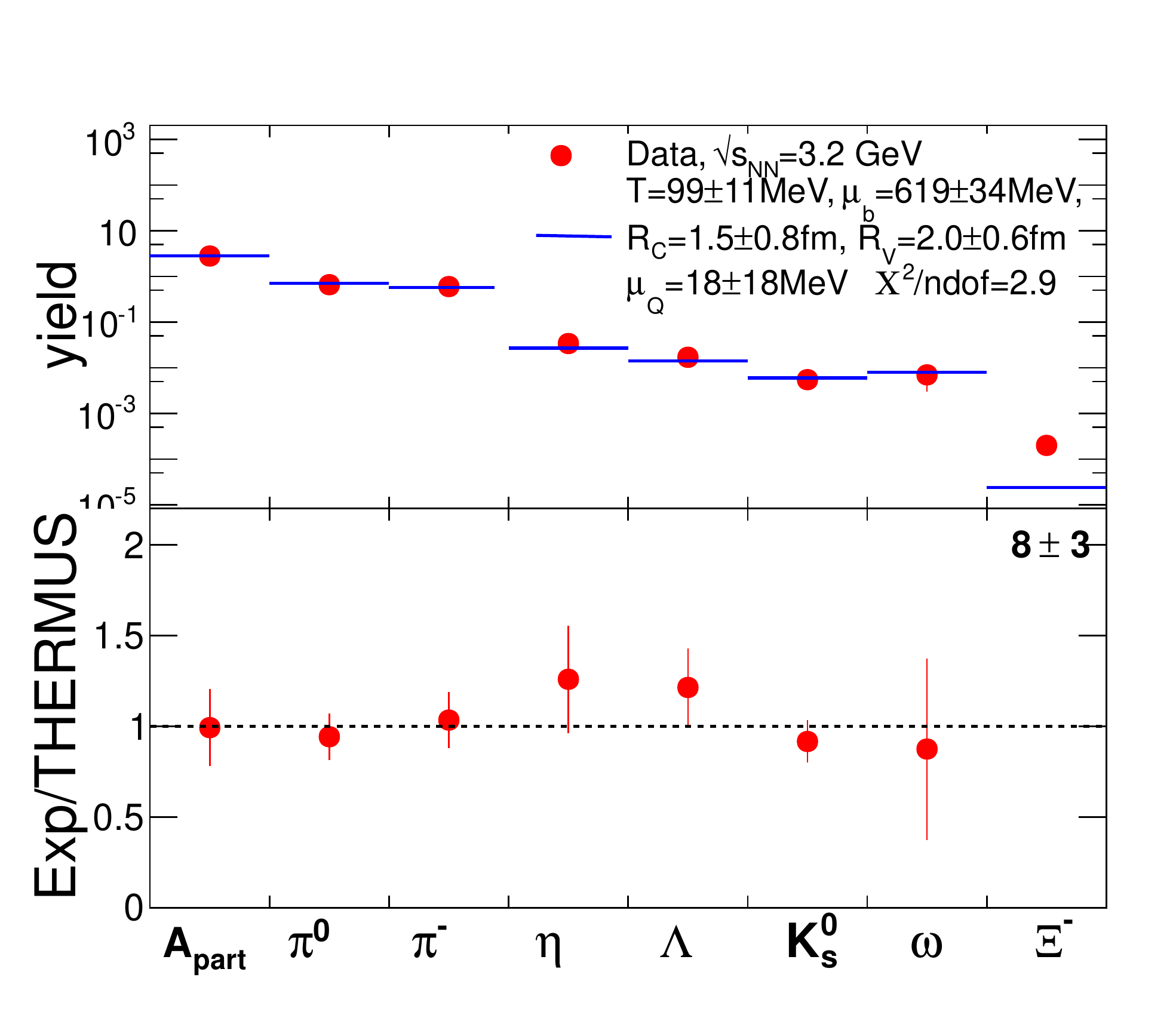}
% epja impact_parameter_herb.eps!!
}
\end{center}
%\vspace{5cm}       % Give the correct figure height in cm
\caption{Yields of hadrons in p+Nb reactions (filled red circles) and the corresponding THERMUS fit (blue bars). The lower plot shows the ratio of the experimental value and the THERMUS value. For the $\Xi^{-}$ the ratio number is quoted instead of displaying a point.}
\label{fig_pNb}       % Give a unique label
\end{figure}
\begin{figure}
\resizebox{8cm}{!}{%
  \includegraphics{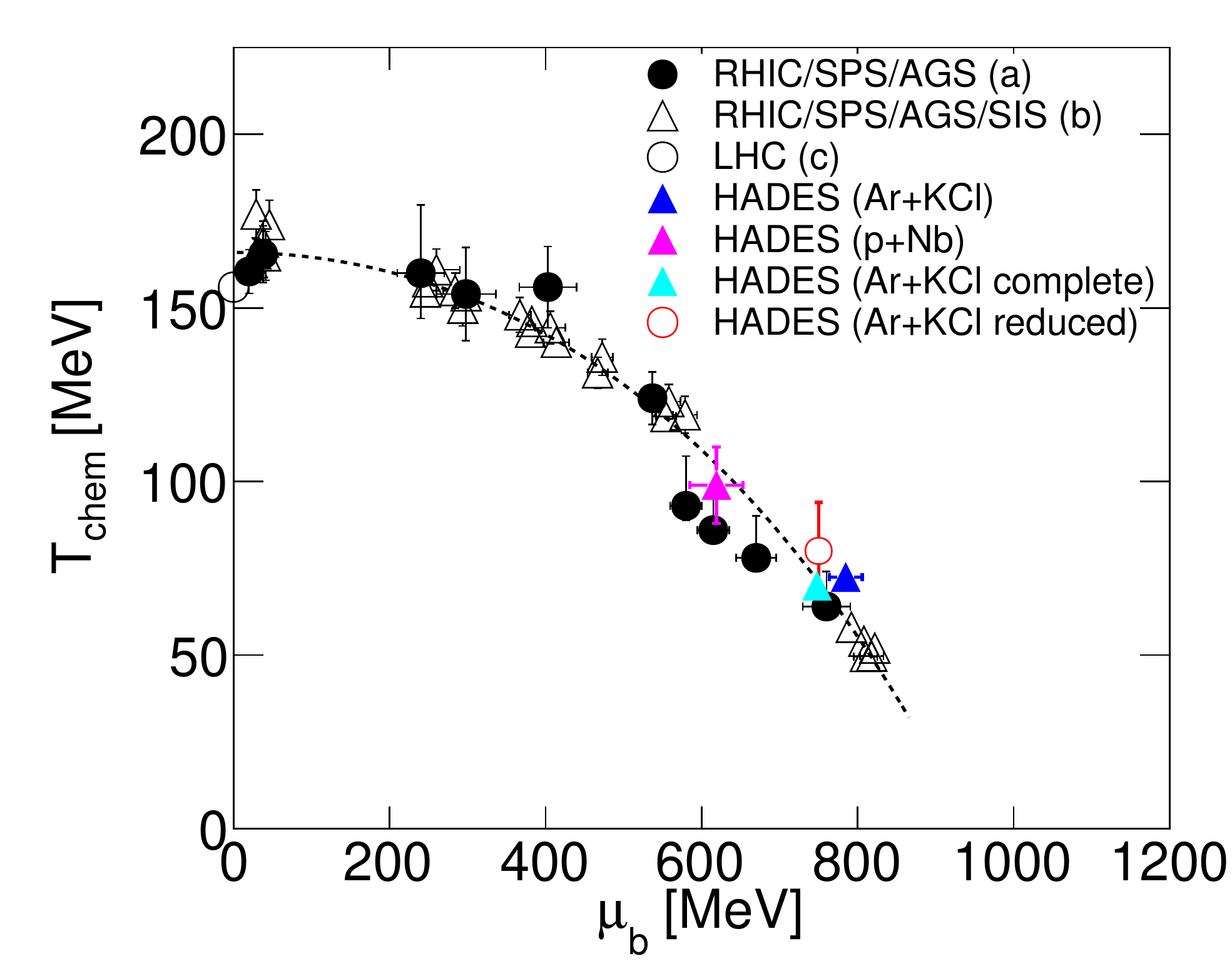}
}
\caption{Chemical freeze-out points in the $T_chem$ --
    $\mu_{b}$ plane. The filled black circles (a) are taken from
    \cite{Andronic:2005yp}, the black open triangles (b) are
    from \cite{Cleymans:2005xv} and the black open circle (c) from \cite{Stachel2013}.
    The presented fit results are defined in the legend.
    The dashed curve correponds to a fixed energy per nucleon of 1 GeV,
    calculated according to \cite{Cleymans:2005xv}. The dark blue triangle corresponds to our Ar+KCl fit presented in \cite{Agakishiev:2010rs}, the light blue triangle shows the result obtained for the full sample and open circle to the reduced data sample both presented in this work.}
\label{fig_TmuB}       % Give a unique label
\end{figure}

For the fit to the yields obtained from p+Nb reactions we add the charge chemical potential $\mu_{Q}$ as an additional free parameter due to the strong asymmetry of the collision system. 
Apart from the charge chemical potential $\mu_{Q}$, we use the same parameters as above. 
The extracted parameters are $T_{chem}=(99\pm11)$ MeV, $\mu_{b}=(619\pm34)$ MeV, $\mu_{Q}=(18\pm18)$ MeV, $R_{c}=(1.5\pm0.8)$ fm, $R=(2.0\pm0.6)$ fm and $\chi^2$/d.o.f. of 2.9.
A detailed comparison of the data with the statistical model fit is shown in the upper part of Fig. \ref{fig_pNb}, while the lower part of this figure depicts the ratio of data to THERMUS values. Again, in case of the $\Xi^-$, a number is displayed instead of a point. \\
Within errors, the values for $R$ and $R_c$ agree with each other, which one expects as the suppression of strange particles compared to non strange particles depends mostly on the absolute value of $R_c$ and only very weakly on the ratio of $R_c$/$R$. \\
The ratio between data and model show striking similarities when comparing the Ar+KCl values in Fig. \ref{fig_arkcl} with the ones of p+Nb in Fig. \ref{fig_pNb}.
In both cases the model is able to describe with fair agreement most of the yields but fails by nearly an order of magnitude in case of the $\Xi^-$.\\
Similar as for the Ar+KCl fit the excess of the experimentally measured $\Xi^-$ yield over the model decreases from a factor $20\pm9$ as reported in \cite{xi_pNb}  to a factor $8\pm3$ when using the current THERMUS version (v3.0).\\
The comparison of the extracted chemical freeze-out temperature $T_{chem}$ to the ones extracted from the inverse slopes of transverse mass spectra at mid-rapidity  for various particles listed in Tab. 2 is more straight forward than in case of Ar+KCl, as we expect no collective expansion of the system. Indeed the extracted slopes show no significant dependence on the particle mass with an average value of $\langle T_{kin} \rangle=(91\pm2)$ MeV, which is in agreement with the value for the chemical freeze-out temperature extracted from the statistical fit of $T_{chem}=(99\pm11)$ MeV.\\
\begin{figure}
\begin{center}
\resizebox{9cm}{!}{%
 \includegraphics{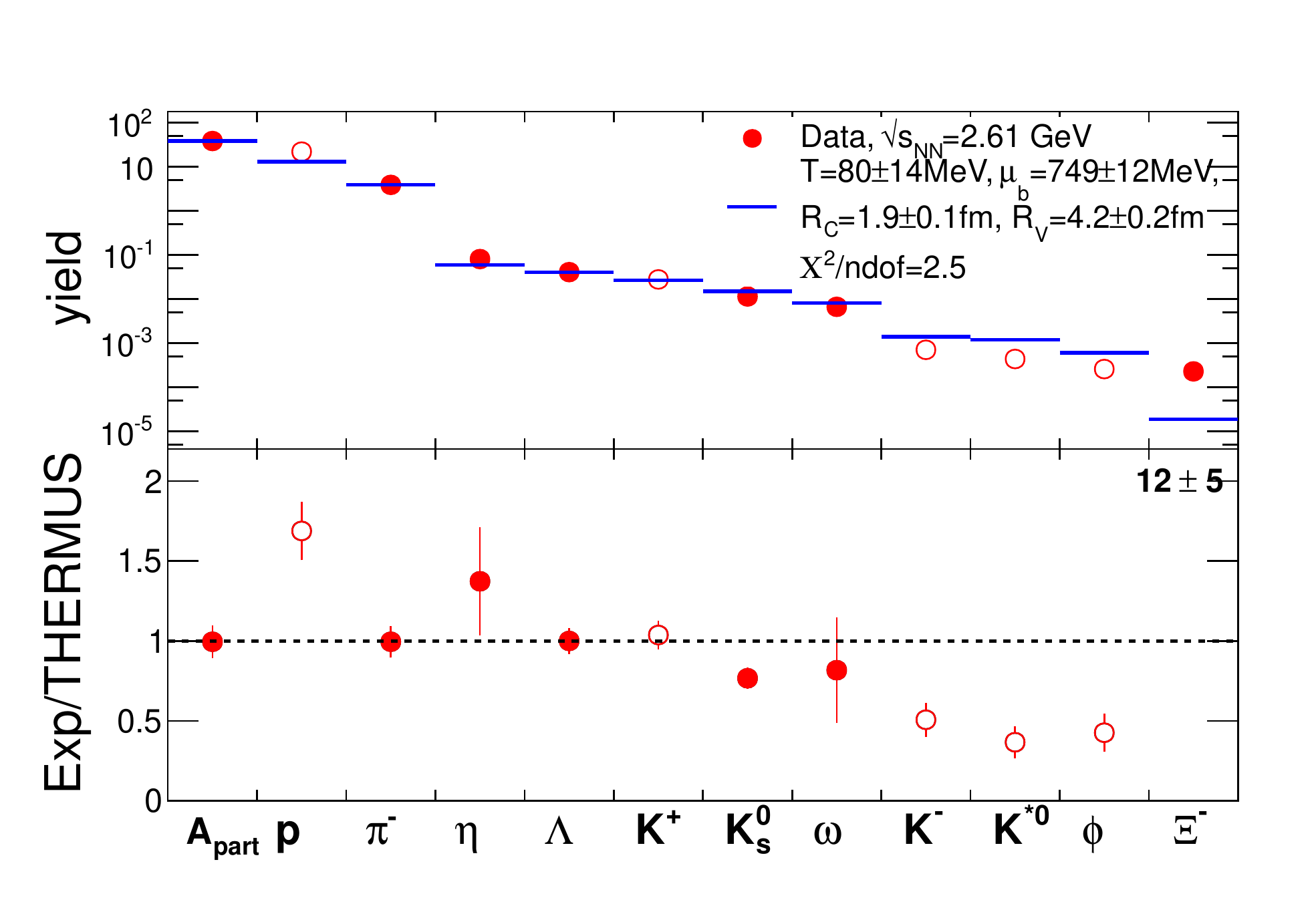}
% epja impact_parameter_herb.eps!!
}
\end{center}
%\vspace{5cm}       % Give the correct figure height in cm
\caption{As in Fig. 1 but excluding the $p$, $K^+$, $K^-$, $K^{*}(892)^0$, $\phi$ yields, shown by open symbols, in the fit.}
\label{fig_pNbalaArKCL}       % Give a unique label
\end{figure}
In order to further discuss the similarities between the Ar+KCl and the p+Nb fit, we reduce our larger Ar+KCl data sample to the same set of identified hadron species as in the p+Nb sample, by excluding the yields of the $p$, $K^+$, $K^-$, $K^{*}(892)^0$ and the $\phi$ in our fit. \footnote{As the yield of the neutral pions is not directly available, we restrict $\mu_{Q}$ by the initial neutron to proton ratio.} 
For the reduced Ar+KCl fit we find $T_{chem}=(80\pm14)$ MeV, $\mu_{b}=(749\pm12)$ MeV, $R_{c}=(1.9\pm0.1)$ fm, $R=(4.2\pm0.2)$ fm and $\chi^2$/d.o.f. of 2.5. The comparison between data and model is shown in a similar way as above in Fig. \ref{fig_pNbalaArKCL}.  \\
By restricting the Ar+KCl sample to a comparable one as available for p+Nb we find a variation of the freeze-out parameters of order 5$\%$, which we attribute to the systematic uncertainties of such an analysis.\\
The $\chi^2$/d.o.f. of the Ar+KCl fits of 3.6 and 2.5 are comparable to the one obtained for the p+Nb sample of 2.9. This is rather surprising as one naively expects a larger amount of thermalization in the larger Ar+KCl system and hence less deviation from statistical equilibrium values. Note that the average number of participants is smaller than 3 in case of the p+Nb sample.  
Furthermore, the p+Nb freeze-out point fits at least as well as the Ar+KCl points to the previously observed regularity of freeze-out points in the  $T_{chem}-\mu_{b}$ plane, displayed in Fig. \ref{fig_TmuB}, where the extracted points of this work are displayed together with similar points extracted in  \cite{Cleymans:2005xv,Andronic:2005yp,Stachel2013}. \\
This brings us back to the motivation of our analysis stressed in the introduction: While the success of the statistical model in describing particle rates from heavy-ion collisions is often implicitly connected to a thermalization of the created system, the success of the model for the p+Nb data  at  $\sqrt{s_{NN}}$= 3.2 GeV questions this connection.\\
Apart from such conceptional issues we stress that $T_{chem}$ for p+Nb at $\sqrt{s_{NN}}$= 3.2 GeV is \textquotedblleft naturally\textquotedblright          ~somewhat larger than $T_{chem}$ for Ar+KCl at $\sqrt{s_{NN}}$= 2.6 GeV which one could attribute to the higher energies in first-chance collisions producing secondary hadrons.  Analogously, $\mu_{b}$ in Ar+KCl is larger than for p+Nb, since some noticeable compression is expected in heavy-ion collisions. On the other hand, the authors of \cite{bec} find only a very small dependence on the system size of $T_{chem}$ and $\mu_{b}$ for SPS energies and hence conclude that both parameters are mainly determined by the energy of the incoming projectile.

\subsection{On the \bf$\Xi^-$ puzzle}
The presence of the excess of the  $\Xi^-$ also in cold nuclear matter has several interesting implications for the interpretation of the heavy-ion data as its origin seems to be already in the elementary channels without the involvement of many-body effects in the medium. Therefore, the increased cross sections of strangeness exchange reactions, which were found to be sufficient to explain the high yield in \cite{Li:2012bga,Graef:2014mra}, seem to be questionable as they are unlikely to play an important role in p+Nb reactions. Also the invoked \cite{Kolomeitsev:2012ij} catalytic strangeness production by secondary processes, such as $\pi + Y \rightarrow \Xi + K$, are strongly suppressed in cold nuclear matter.\\
Feed down from the high-mass tails of resonances has been proposed as another possible explanation recently in \cite{Steinheimer:2015sha}.  The authors tuned the mass depending branching ratio of high-lying baryon resonances, namely the $N^{*}(1990)$, $N^{*}(2080)$, $N^{*}(2190)$, $N^{*}(2220)$ and $N^{*}(2250)$, in a transport code to match elementary data on $\phi$ meson production. As a result the $\phi$/$K^-$ ratio in Ar+KCl is successfully reproduced. The same mechanism is then also used for the  $\Xi^-$ hyperon but due to the lack of elementary data the model is tuned to match our p+Nb data. \\
The technique of mass dependent branching ratios for broad resonances has been successfully applied in order to describe the dilepton spectra at low energies as pointed out in \cite{Weil:2012ji,Brat}. Although the tuned branching ratios are still consistent with the OZI rule there is no experimental evidence for the decay of the $N^{*}$ resonance to final states containing a $\phi$ meson or a $\Xi$ hyperon. Therefore, there is no branching of $N^{*}$ resonances to double-strange final states included in THERMUS. \\
In addition, the total yield of the five above discussed $N^{*}$ resonances amounts to $1.47 \times 10^{-3}$ and $7.1 \times 10^{-4}$ in THERMUS for the Ar+KCl and the p+Nb data sample respectively. In order to explain the observed $\Xi^-$ yields which are at the order of $10^{-4}$ in both systems, branching ratios of the $N^{*}$ resonances to $\Xi^-$ at the order of $10\%$ or higher would be needed. This seems to be unreasonably high. \\
Instead, the feed down to $\Xi$ hyperons is mainly originating from the decay of excited $\Xi$(1530,1690,1820,1950,2030) and some higher-lying $\Lambda$ and $\Sigma$ states. Note that the widths, branching ratios etc. of these states are not well constrained and subject of current and future scientific activities \cite{ron,Lutz}. 
The feed down fraction to the  $\Xi^-$ yield amounts to 10$\%$ and 27$\%$ percent of the total rate in THERMUS respectively for Ar+KCl and p+Nb. From these numbers it becomes clear that the difference in yields of the double-strange hyperons of about $50\%$ between the two used THERMUS versions is not due to additional higher-lying states and corresponding feed-down.
Instead, the difference is connected to the new states in the hadron spectrum via the mechanism for exact strangeness conservation using the strangeness correlation volume $V_c$. As there are more hadronic (mesonic) states included in the current version, the probability for counterbalancing the strange quarks of the hyperons is higher for a given volume. Especially the rather low-lying scalar meson $K^{*}(800)^{0+}$ (also known as $\bf\kappa$) is important in this context as its abundance is still not too rare. However, one should note that, with a widths of $\Gamma \approx 500$ MeV evidences for this state are hard to establish.
The effect on the yield is much more prominently seen in the double-strange baryon sector because their yield scales with the volume proportional to the power of three, see Eq. (6) compared to the one of single-strange particles whose yield scales only quadratically with the volume, see Eq. (5).\\
From this consideration it becomes clear that a precise knowledge of the hadron spectrum is an important issue for the interpretation of HIC data.

\section{Summary and Outlook}

By comparing the obtained freeze-out parameters from a statistical model fit to HADES data obtained from p+Nb and Ar+KCl collisions at center of mass energies of $\sqrt{s_{NN}}$ = 3.2 GeV and  $\sqrt{s_{NN}}=$ 2.6 GeV respectively,  
we tried to address the aspect of equilibration in HIC by testing two manifestations, the description of yields and regularity of freeze-out parameters. We make the rather surprising finding that the statistical model is able to describe the p+Nb data as well as the larger system the Ar+KCl data, which questions the often drawn connection between the agreement of statistical models with particle yields in heavy-ion collisions and thermalization.\\
Furthermore, we emphasize that the excess of the $\Xi^-$ is already present in cold nuclear matter. Given the rates of higher-lying $N^{*}$ resonances predicted by our statistical model fit, we find feed down of these states a rather implausible explanation for the excess of the $\Xi^-$ yield over the model value. In addition, we state the importance of a precise knowledge of the hadron spectrum for interpretation of HIC data.\\
We want to point out that HADES data of central Au+Au collisions will be available soon and  might allow to gather further insights into the subject. \\
\\
{\bf Acknowledgments }\\
LIP Coimbra, Coimbra (Portugal) PTDC/FIS/113339/ 2009 SIP JUC Cracow, Cracow (Poland) NCN grant 2013/ 10/M/ ST2/00042, N N202 286038 28-JAN-2010 NN202198639 01-OCT-2010 Helmholtz-Zentrum Dresden-Rossendorf (HZDR), Dresden (Germany) BMBF 06DR9059D \\   
TU M\"unchen, Garching (Germany) MLL M\"unchen DFG EClust 153 VH-NG-330 BMBF 06MT9156 TP5 GSI\\
TMKrue 1012 NPI AS CR, Rez, Rez (Czech Republic) MSMT LC07050 GAASCR IAA100480803 USC - S. de Compostela, Santiago de Compostela (Spain) CPAN:\\
CSD2007-00042 Goethe-University, Frankfurt (Germany) HA216/EMMI HIC for FAIR (LOEWE) BMBF: 06FY9100I GSI FE EU Contract No. HP3-283286. One of us (M.L.) acknowledges the support of the Humboldt Foundation.

\newpage
\appendix
\section{Statistical strangeness production at high baryo-chemical potential using exact strangeness conservation}
In a pure statistical ansatz based on the common temperature $T$ of all species and the baryo-chemical potential $\mu_{B}$, the multiplicities of mesons and baryons produced in a heavy-ion collision, neglecting feed-down and isospin asymmetry are given by
\begin{equation}
\sum_{i} M_{m_{i}}=\sum_{i}g_{i}V\int \frac{d^{3}p}{(2\pi)^{3}}\exp\left(-\frac{E_{i}}{T}\right) \times F_{Si},
\label{1_eq}
\end{equation}
\begin{equation}
\sum_{j} M_{b_{j}}=\sum_{j}g_{j}V\int \frac{d^{3}p}{(2\pi)^{3}}\exp\left(-\frac{E_{j}-\mu_{B}}{T}\right) \times F_{Sj},
\label{2_eq}
\end{equation}
with $M_{m_{i}}$ and $M_{b_{j}}$ being the multiplicities of a given meson (m) or baryon (b), the degeneracy factors $g_{i,j}$, the volume of the fireball V and the energies of the corresponding meson and baryon $E_{i,j}=\sqrt{m_{i,j}^{2}+p_{i,j}^{2}}$. \\ 
Particles containing strangeness are rare, especially at SIS energies, and therefore the strange quantum number must be exactly conserved in each event in the ensemble; each particle carrying a strange quark must be counterbalanced by one carrying an antistrange quark due to associated strangeness production in strong-interaction processes. 
This results in a multiplicative canonical suppression factor \\
$F_{S}(T,\mu_{B},V,S,N_{S})$, which is equal to one for non-strange particles and approaches, in the limit of large volumes and temperatures, the grand canonical fugacities \\ $\lim\limits_{V,T \rightarrow \infty}{F_{Si,j}}=exp\left(-S_{i,j}\mu_{S}/T\right)$. \ 
The factor $F_{Si,j}$ for each particle species $i,j$ depends in general on the thermodynamical properties of the system, the strangeness content $S_{i,j}$ of the respective particle and the number of meson and baryon states containing strangeness $N_{S}$.\\ 
We illustrate the effect of $F_{S}$ by making use of the opposite limit of small volume and temperature, where particle numbers are small and the canonical strangeness suppression is most relevant.  For $M<1$ and neglecting higher order effects, the multiplicity of kaons $M_{m_{K}}$ can be approximated as
\begin{align}
M_{m_{K}}\approx g_{K}V\int \frac{d^{3}p}{(2\pi)^{3}}\exp\left(-\frac{E_{K}}{T}\right) \times  \left[ g_{Y}V\int \frac{d^{3}p}{(2\pi)^{3}} \right. \nonumber\\
\left. \exp\left(-\frac{E_{Y}-\mu_{B}}{T}\right)  +  g_{\overline{K}}V\int \frac{d^{3}p}{(2\pi)^{3}}\exp\left(-\frac{E_{\overline{K}}}{T}\right)  \right] .
\label{3_eq}
\end{align}
The two terms inside the square brackets correspond to the counterbalance terms for the antistrange quark inside the kaons from the hyperons and from the antikaons.
Hence Eq. (3) can be rewritten as

\begin{equation}
M_{m_{K}} \approx M^{GC}_{m_{K}} \times \left[ M^{GC}_{m_{Y}} + M^{GC}_{m_{\overline{K}}} \right]
\label{4_eq}
\end{equation}
with $M^{GC}_{b_{Y}}, M^{GC}_{m_{K}}$ and $M^{GC}_{m_{\overline{K}}}$ corresponding to the grand-canonical multiplicities of kaons, hyperons and antikaons, respectively.
As we assume $T$ and $V$ to be sufficiently small, so that $M<1$, one can clearly see the resulting suppression due to exact strangeness conservation.
Due to the absence of antimatter at high $\mu_{B}$ the counterbalance term for the antikaons by the antihyperons is missing, resulting in a stronger suppression compared to kaons. Hence for $(E_{Y}-\mu_{B}) \ll E_{K}$ we can neglect the antikaon term and approximate the  multiplicity of single-strange hyperons as
\begin{equation}
M_{b_{Y}} \approx M^{GC}_{b_{Y}} \times \left[ M^{GC}_{m_{K}}  \right]\propto V^2.
\label{4_eq}
\end{equation}
Compared to non-strange particle multiplicities, which are proportional to $V$ the multiplicity of single-strange particles is proportional to $V^2$ for small systems.
Rewriting Eq. (5) for double-strange hyperons, like the $\Xi$, one finds their yield scaling with $V^3$:
\begin{equation}
M_{b_{\Xi}} \approx M^{GC}_{b_{\Xi}} \times \left[ M^{GC}_{m_{K}} \right]^{2}\propto V^3.
\label{5_eq}
\end{equation}
The above mentioned strangeness correlation volume $V_c$, in which strangeness has to be exactly conserved, is realized by setting the volume terms in the brackets in Eq. (3) to $V_c$. For $V_{c}<V$, strange particles are suppressed additionally on top of the pure canonical suppression. \\ 
From these considerations it becomes clear that the size of $F_{S}$ for a given volume $V_c$ and hence the strength of the suppression depends also on the number of known strange particle states. The more states exist the more possibilities are available for counterbalancing the strange (anti-strange) quarks. This is of relevance for the comparison of the different THERMUS versions 2.3 and 3.0. The latter one has more states included, especially the rather low lying  K*(800) states are important in this context. 


\begin{thebibliography}{9}
	\bibitem{koppe}
	H. Koppe, Z. Naturforschg. 3 a, 251 (1948).
	
	\bibitem{tawfik}
	A.~N.~Tawfik, Z. Naturforschg. 69a, 106 (2014).
	
	\bibitem{BraunMunzinger:2003zd}
	P.~Braun-Munzinger, K.~Redlich and J.~Stachel,
	%``Particle production in heavy ion collisions,''
	arXiv:nucl-th/0304013 (2003).
	%%CITATION = NUCL-TH/0304013;%%
	%\cite{Cleymans:2005xv}
	
	\bibitem{Tawfik:2014eba}
	A.~N.~Tawfik,
	%``Equilibrium statistical-thermal models in high-energy physics,''
	Int.\ J.\ Mod.\ Phys.\ A {\bf 29},  1430021 (2014).
	
	\bibitem{Floris}
	M.~Floris,
	%``Hadron yields and the phase diagram of strongly interacting matter,''
	Nucl.\ Phys.\ A {\bf 931}, 103 (2014).
	
	\bibitem{Cleymans:2005xv}
	J.~Cleymans, H.~Oeschler, K.~Redlich and S.~Wheaton,
	%``Comparison of chemical freeze-out criteria in heavy-ion collisions,''
	Phys.\ Rev.\  C \textbf{73}, 034905 (2006).
	
	\bibitem{hage}
	R.~Hagedorn,
	%``Statistical thermodynamics of strong interactions at high-energies,''
	Nuovo Cim.\ Suppl.\  {\bf 3}, 147 (1965).
	
	\bibitem{shuryak}
	E.~V.~Shuryak,
	%``Multiparticle production in high energy particle collisions.,''
	Yad.\ Fiz.\  {\bf 16}, 395 (1972).
	
	\bibitem{Becattini:2003wp}
	F.~Becattini, M.~Gazdzicki, A.~Keranen, J.~Manninen and R.~Stock,
	%``Study of chemical equilibrium in nucleus nucleus collisions at AGS and  SPS
	%energies,''
	Phys.\ Rev.\  C \textbf{69}, 024905 (2004).  
	
	\bibitem{Becattini:2001fg}
	F.~Becattini and G.~Passaleva,
	%``Statistical hadronization model and transverse momentum spectra of hadrons in high-energy collisions,''
	Eur.\ Phys.\ J.\ C {\bf 23}, 551  (2002).
	%\cite{Becattini:2008tx}
	
	\bibitem{Becattini:2008tx}
	F.~Becattini, P.~Castorina, J.~Manninen and H.~Satz,
	%``The Thermal Production of Strange and Non-Strange Hadrons in e+ e- Collisions,''
	Eur.\ Phys.\ J.\ C {\bf 56}, 493 (2008).
	
	\bibitem{andronic_el}
	A.~Andronic, F.~Beutler, P.~Braun-Munzinger, K.~Redlich and J.~Stachel,
	%``Thermal description of hadron production in e+e- collisions revisited,''
	Phys.\ Lett.\ B {\bf 675}, 312 (2009).
	
	\bibitem{Andronic:2005yp}
	A.~Andronic, P.~Braun-Munzinger and J.~Stachel,
	%``Hadron production in central nucleus nucleus collisions at chemical
	%freeze-out,''
	Nucl.\ Phys.\  A \textbf{772}, 167 (2006).
	
	\bibitem{Stachel2013}
	J.~Stachel, A.~Andronic, P.~Braun-Munzinger and K.~Redlich,
	%``Confronting LHC data with the statistical hadronization model,''
	J.\ Phys.\ Conf.\ Ser.\  {\bf 509}, 012019 (2014).
	
	%becattini heavy/ion
	
	\bibitem{Cleymans:1999} J.~Cleymans {\it et al.}, Phys.\ Rev.\  C {\bf 59}, 1663 (1999).
	%\cite{Andronic:2009gj}
	
	\bibitem{Kraus:2007hf}
	I.~Kraus, J.~Cleymans, H.~Oeschler, K.~Redlich and S.~Wheaton,
	%``Chemical Equilibrium in Collisions of Small Systems,''
	Phys.\ Rev.\  C \textbf{76}, 064903 (2007).
	
	\bibitem{Petran}
	M.~Petran, J.~Letessier, V.~Petracek and J.~Rafelski,
	%``Hadron production and quark-gluon plasma hadronization in Pb-Pb collisions at $\sqrt{s_{NN}}=2.76$ TeV,''
	Phys.\ Rev.\ C {\bf 88},  034907 (2013).
	
	%\cite{Cleymans:2003yp}
	\bibitem{cley1}
	J.~Cleymans, B.~K\"ampfer, P.~Steinberg and S.~Wheaton,
	%``System size dependence of strangeness saturation,''
	J.\ Phys.\ G {\bf 30},  S595 (2004).
	%[hep-ph/0311020].
	%%CITATION = HEP-PH/0311020;%%
	%8 citations counted in INSPIRE as of 22 sept. 2015
	
	\bibitem{cley2}
	B.~K\"ampfer, J.~Cleymans, P.~Steinberg and S.~Wheaton,
	%``Strangeness saturation: Dependence on system size, centrality and energy,''
	Heavy Ion Phys.\  {\bf 21}, 207 (2004). 
	%[hep-ph/0304269].
	%%CITATION = HEP-PH/0304269;%%
	%5 citations counted in INSPIRE as of 22 sept. 2015
	
	%\cite{Cleymans:2004pp}
	\bibitem{bk3}
	J.~Cleymans, B.~Kampfer, M.~Kaneta, S.~Wheaton and N.~Xu,
	%``Centrality dependence of thermal parameters deduced from hadron multiplicities in Au + Au collisions at s(NN)**(1/2) = 130-GeV,''
	Phys.\ Rev.\ C {\bf 71}, 054901 (2005). 
	%doi:10.1103/PhysRevC.71.054901
	%[hep-ph/0409071].
	%%CITATION = doi:10.1103/PhysRevC.71.054901;%%
	%91 citations counted in INSPIRE as of 25 Nov 2015
	
	\bibitem{Agakishiev:2010rs}
	G.~Agakishiev {\it et al.}  [HADES Collaboration],
	%``Hyperon production in Ar+KCl collisions at 1.76A GeV,''
	Eur.\ Phys.\ J.\ A {\bf 47}, 21 (2011).
	
	% %\cite{Steinheimer:2015sha}
	\bibitem{Steinheimer:2015sha}
	J.~Steinheimer and M.~Bleicher,
	%``Sub-threshold $\phi$ and $\Xi^-$ production by high mass resonances with UrQMD,''
	arXiv:1503.07305 [nucl-th].
	%%CITATION = ARXIV:1503.07305;%%
	
	\bibitem{Wheaton:2004qb}
	S.~Wheaton and J.~Cleymans,
	%``THERMUS: A thermal model package for ROOT,''
	Comput.\ Phys.\ Commun.\ \textbf{180}, 84 (2009).
	
	%\cite{Agashe:2014kda}
	\bibitem{Agashe:2014kda} 
	K.~A.~Olive {\it et al.} [Particle Data Group Collaboration],
	%``Review of Particle Physics,''
	Chin.\ Phys.\ C {\bf 38}, 090001 (2014).
	%%CITATION = CHPHD,C38,090001;%%
	%1811 citations counted in INSPIRE as of 08 sept. 2015
	
	%include in elementary discussion
	%\cite{Pulawski:2015tka}
	%\bibitem{Pulawski:2015tka}
	%  S.~Puławski [NA61 Collaboration],
	%``Energy dependence of hadron spectra and multiplicities in p+p interactions,''
	%  arXiv:1502.07916 [nucl-ex].
	%%CITATION = ARXIV:1502.07916;%%
	
	%%HADES paper
	\bibitem{Agakishiev:2009am}
	G.~Agakishiev {\it et al.}  [HADES Collaboration],
	%``The High-Acceptance Dielectron Spectrometer HADES,''
	Eur.\ Phys.\ J.\  A \textbf{41}, 243 (2009).
	
	\bibitem{schuldes_fair}
	H.~Schuldes {\it et al.}  [HADES Collaboration],
	%``Protons and light fragments in Ar+KCl at 1.76 AGeV measured with HADES,''
	J.\ Phys.\ Conf.\ Ser.\  {\bf 599},  012028 (2015).
		
	%\cite{Tlusty:2009dk}
	\bibitem{Pavel}
	P.~Tlusty {\it et al.} [HADES Collaboration],
	%``Charged pion production in C+C and Ar+KCl collisions measured with HADES,''
	arXiv:0906.2309 (2009).
	%%CITATION = ARXIV:0906.2309;%%
	%13 citations counted in INSPIRE as of 20 Nov 2015
	
    \bibitem{Agakishiev:2009ar}
	G.~Agakishiev {\it et al.}  [HADES Collaboration],
	%``Phi decay: a relevant source for K- production at SIS energies?,''
	Phys.\ Rev.\  C \textbf{80}, 025209 (2009).
	
	\bibitem{Agakishiev:2010zw} 
	G.~Agakishiev {\it et al.} [HADES Collaboration], Phys. Rev. C \textbf{82}, 044907 (2010).
	%``In-Medium Effects on K$^0$ Mesons in Relativistic Heavy-Ion Collisions,''
	%\cite{Agakishiev:2009rr} 
		
	\bibitem{Agakishiev:2013nta}
	G.~Agakishiev {\it et al.}  [HADES Collaboration],
	%``Deep sub-threshold $K^{*}(892)^{0}$ production in collisions of Ar + KCl at 1.76-A-GeV,''
	Eur.\ Phys.\ J.\ A {\bf 49}, 34  (2013).
	
	\bibitem{Agakishiev:2011vf}
	G.~Agakishiev {\it et al.}  [HADES Collaboration],
	%``Dielectron production in Ar+KCl collisions at 1.76A GeV,''
	Phys.\ Rev.\ C {\bf 84},  014902 (2011).
	%\cite{Agakishiev:2013nta}
	
	\bibitem{Agakishiev:2009rr}
	G.~Agakishiev {\it et al.}  [HADES Collaboration],
	%``Deep sub-threshold $\Xi^-$ production in Ar+KCl reactions at 1.76A GeV,''
	Phys.\ Rev.\ Lett.\  \textbf{103}, 132301 (2009).
	
	\bibitem{UrQMD} S.\,A. Bass {\it et al.}, Prog. Part. Nucl. Phys. \textbf{41}, 225 (1998).
	
	\bibitem{Averbeck:2000sn}
	R.~Averbeck, R.~Holzmann, V.~Metag and R.~S.~Simon,
	%``Neutral pions and eta mesons as probes of the hadronic fireball in  nucleus
	%nucleus collisions around 1-A-GeV,''
	Phys.\ Rev.\  C \textbf{67}, 024903 (2003).
	
	\bibitem{Agakishiev:2013noa}
	G.~Agakishiev {\it et al.}  [HADES Collaboration],
	%``Inclusive pion and η production in p+Nb collisions at 3.5 GeV beam energy,''
	Phys.\ Rev.\ C {\bf 88},  024904 (2013).
	%\cite{Agakishiev:2014kdy}
	\bibitem{Agakishiev:2014kdy}
	G.~Agakishiev {\it et al.}  [HADES Collaboration],
	%``Lambda hyperon production and polarization in collisions of p(3.5 GeV)+Nb,''
	Eur.\ Phys.\ J.\ A {\bf 50},  81 (2014).
	%\cite{Agakishiev:2014moo}
	\bibitem{Agakishiev:2014moo}
	G.~Agakishiev {\it et al.}  [HADES Collaboration],
	%``Medium effects in proton-induced $K^{0}$ production at 3.5 GeV,''
	Phys.\ Rev.\ C {\bf 90}, 054906 (2014).
	\bibitem{xi_pNb}
	G.~Agakishiev {\it et al.}  [HADES Collaboration].
	%``Subthreshold Xi- Production in Collisions of p(3.5 GeV)+Nb,''
	Phys.\ Rev.\ Lett.\  \textbf{114}, 212301 (2015).
	%%CITATION = ARXIV:1501.03894;%%
	%2 citations counted in INSPIRE as of 19 Apr 2015
	%
	\bibitem{Agakishiev:2012vj}
	G.~Agakishiev {\it et al.}  [HADES Collaboration],
	%``First measurement of low momentum dielectrons radiated off cold nuclear matter,''
	Phys.\ Lett.\ B {\bf 715},  304 (2012).
	%\cite{Agakishiev:2013noa}
	%\cite{Weil:2012ji}
	\bibitem{glauber}
	R. J. Glauber and G. Matthiae,
	Nucl. Phys.  \textbf{B 21}, 135 (1970).
	
	\bibitem{Weil:2012ji}
	J.~Weil, H.~van Hees and U.~Mosel,
	%``Dilepton production in proton-induced reactions at SIS energies with the GiBUU transport model,''
	Eur.\ Phys.\ J.\ A {\bf 48},  111 (2012).
	\bibitem{Weil_priv}
	J.~Weil, private communication.
	
	\bibitem{pdg2002}
	K. Hagiwara et al. (Particle Data Group), Phys. Rev. D \textbf{66}, 010001 (2002).
	
	%\cite{Markert:2002rw}
	\bibitem{Markert:2002rw}
	C.~Markert, G.~Torrieri and J.~Rafelski,
	%``Strange hadron resonances: Freezeout probes in heavy ion collisions,''
	AIP Conf.\ Proc.\  {\bf 631}, 533 (2002).
	%[hep-ph/0206260].
	%%CITATION = HEP-PH/0206260;%%
	%37 citations counted in INSPIRE as of 24 Apr 2015
	%\cite{Kolomeitsev:2012ij}
	
	%\cite{Becattini:2005xt}
	\bibitem{bec}
	F.~Becattini, J.~Manninen and M.~Gazdzicki,
	%``Energy and system size dependence of chemical freeze-out in relativistic nuclear collisions,''
	Phys.\ Rev.\ C {\bf 73}, 044905 (2006).
	%[hep-ph/0511092].
	%%CITATION = HEP-PH/0511092;%%
	%271 citations counted in INSPIRE as of 24 sept. 2015
	%\cite{Stachel:2013zma}
		
	%\cite{Li:2012bga}
	\bibitem{Li:2012bga}
	F.~Li, L.~W.~Chen, C.~M.~Ko and S.~H.~Lee,
	%``Contributions of hyperon-hyperon scattering to subthreshold cascade production in heavy ion collisions,''
	Phys.\ Rev.\ C {\bf 85}, 064902 (2012).
	%[arXiv:1204.1327 [nucl-th]].
	%%CITATION = ARXIV:1204.1327;%%
	%7 citations counted in INSPIRE as of 24 Apr 2015
	%  
	%%\cite{Graef:2014mra}
	\bibitem{Graef:2014mra}
	G.~Graef, J.~Steinheimer, F.~Li and M.~Bleicher,
	%``Deep sub-threshold $\Xi$ and $\Lambda$ production in nuclear collisions with the UrQMD transport model,''
	Phys.\ Rev.\ C {\bf 90},  064909 (2014).
	%[arXiv:1409.7954 [nucl-th]].
	%%CITATION = ARXIV:1409.7954;%%
	%3 citations counted in INSPIRE as of 24 Apr 2015
	
	\bibitem{Kolomeitsev:2012ij}
	E.~E.~Kolomeitsev, B.~Tomasik and D.~N.~Voskresensky,
	%``Strangeness Balance in HADES Experiments and the $\Xi^-$ Enhancement,''
	Phys.\ Rev.\ C {\bf 86}, 054909 (2012) .
	%[arXiv:1207.5738 [nucl-th]].
	%%CITATION = ARXIV:1207.5738;%%
	%5 citations counted in INSPIRE as of 24 Apr 2015
	
	\bibitem{mB}
	M.~Lorenz {\it et al.} (HADES Collaboration),
	%Contributed to 48th International Winter Meeting on Nuclear Physics, Bormio, Italy, 25-29 Jan 2010.
	PoS (BORMIO2010) 038  (2010).
	
	\bibitem{fphi}
	K.~Piasecki {\it et al.}  [FOPI Collaboration],
	%``Influence of $\phi$ mesons on negative kaons in Ni+Ni collisions at 1.91A GeV beam energy,''
	Phys.\ Rev.\ C {\bf 91},  054904 (2015).
	%%CITATION = ARXIV:1412.4493;%%
	
	%\cite{Cheng:2003as}
	\bibitem{Xu}
	Y.~Cheng, F.~Liu, Z.~Liu, K.~Schweda and N.~Xu,
	%``Transverse expansion in Au-197 + Au-197 collisions at rhic,''
	Phys.\ Rev.\ C {\bf 68},  034910 (2003).
	%%CITATION = PHRVA,C68,034910;%%
	%22 citations counted in INSPIRE as of 24 Apr 2015
	
	%\cite{Bratkovskaya:2013vx}
	\bibitem{Brat}
	E.~L.~Bratkovskaya, J.~Aichelin, M.~Thomere, S.~Vogel and M.~Bleicher,
	%``System size and energy dependence of dilepton production in heavy-ion collisions at 1-2 GeV/nucleon energies,''
	Phys.\ Rev.\ C {\bf 87}, 064907 (2013).
	%[arXiv:1301.0786 [nucl-th]].
	%%CITATION = ARXIV:1301.0786;%%
	%15 citations counted in INSPIRE as of 09 sept. 2015
	
	%\cite{Lutz:2009ff}
	\bibitem{ron}
	M. Ronniger and B. Ch. Metsch,
	Eur.\ Phys.\ J. A  {\bf 47}, 162 (2011).
	
	%\cite{Lutz:2009ff}
	\bibitem{Lutz}
	M.F.M.~Lutz {\it et al.} [PANDA Collaboration],
	%``Physics Performance Report for PANDA: Strong Interaction Studies with Antiprotons,''
	arXiv:0903.3905 (2009).
	%%CITATION = ARXIV:0903.3905;%%
	%315 citations counted in INSPIRE as of 26 Oct 2015
	
	%\cite{Arnaldi:2008fw}
	%\bibitem{NA60}
	%  R.~Arnaldi {\it et al.}  [NA60 Collaboration],
	%``NA60 results on thermal dimuons,''
	%  Eur.\ Phys.\ J.\ C {\bf 61} (2009) 711
	%  [arXiv:0812.3053 [nucl-ex]].
	%%CITATION = ARXIV:0812.3053;%%
	%85 citations counted in INSPIRE as of 24 Apr 2015
	
	
\end{thebibliography}
\end{document}